\documentclass[12pt]{article}
\usepackage[left=1.1in,right=1.1in,top=1.2in,bottom=1.2in,footnotesep=.25in]{geometry}
\usepackage{authblk}

\usepackage{amssymb}
\usepackage{amsmath}
\usepackage{xcolor}
\usepackage{graphicx}
 \usepackage{floatrow}
 \usepackage{soul}
 \usepackage[font=small,labelfont=bf]{caption}
 \usepackage[margin=.8cm]{caption}

\date{}
\title{Cherenkov radiation as ghost instability}

\author{Eugeny~Babichev} 
\affil{Universit\'e Paris-Saclay, CNRS/IN2P3, IJCLab, 91405 Orsay, France}

\begin{document}

\maketitle

\begin{abstract}
We demonstrate that Cherenkov radiation can be interpreted as ghost instability of a certain type. 
Solutions of modified gravity theories often contain ghost instabilities. 
One type of such instability is associated with existence of different types of species with causal cones that do not share common time, which leads to vacuum decay via creation of particles with positive and negative energies. 
We show that this ghost instability can be seen as Cherenkov radiation and vice versa.
\end{abstract}

\section{Introduction}
Cherenkov radiation is a well-known effect, it is normally referred to electromagnetic radiation emitted by a charged particle that passes through a medium with velocity greater than a speed of light in the medium~\cite{Frank:1937fk,ginzburg,jackson,nair}. 
The mechanism of the Cherenkov radiation can be understood in terms of interference of radiation emitted by a moving particle, or as a resonance effect~\cite{ginzburg}.
With these interpretations it is clear that Cherenkov radiation arises for waves of any kind. 
In particular, a shock wave generated by a fast moving aircraft can be also considered as Cherenkov radiation.
From a more general viewpoint, the Cherenkov effect exists for a particle moving faster than the speed of propagation of perturbations in any medium, provided that there is an interaction between the moving particle and the medium.
For simplicity, we will refer to perturbations in a medium as phonons.

In this paper we give yet another point of view on Cherenkov radiation. 
We show that it can be considered as ghost instability. 
Ghost is commonly understood as a state or a particle with negative energy (i.e. we consider physical ghost degrees of freedom, not the extraneous fields introduced for gauge-fixing). 
Therefore if there is an interaction between a ghost particle and normal non-ghost particle (that is one with positive energy), 
one may expect a catastrophic vacuum decay. 
Indeed, if there are particles with both positive and negative energies, the energy conservation does not forbid creation `from nothing' normal and ghost particles, thus making the state with the ghost particles in the spectrum unstable, see e.g.~\cite{Cline:2003gs}. 

Ghosts and other types of instabilities are quite common in modern modifications of gravity, see e.g. Refs.~\cite{Clifton:2011jh,Joyce:2014kja,deRham:2014zqa,Kobayashi:2019hrl} for review on gravity modifications.
Apart from instabilities, there are other pathologies that can accompany new theories, including strong coupling and 
the problem of caustics~\cite{Babichev:2016hys,Babichev:2017lrx,Tanahashi:2017kgn,Reall:2014sla}.
Since normally theories are highly non-linear, the standard approach is to consider linear perturbations on top of a background solution and then to conclude based on the properties of linearised degrees of freedom, whether the solution is stable or not. 
It is then worth mentioning that there are subtleties related to the identification of negative energy (or Hamiltonian) and the question of stability.
Indeed, the Hamiltonian unbounded from below does not necessarily imply instability of a solution, it might be an effect of a bad choice of coordinates~\cite{Babichev:2017lmw,Babichev:2018uiw}.
In this paper  we will be interested in a certain type of a ghost instability that often arises in solutions of modified gravity. 
Our consideration here is however fully general, since we do not make any reference to a specific gravity theory.

Here we show that one can consider two seemingly very different physical effects, Cherenkov radiation and ghost instability, from the same perspective.
I.e. we will demonstrate that one can interpret Cherenkov radiation as instability with creation of ghost carrying negative energy. 
And the other way around, the vacuum decay process due to the presence of a ghost can be seen as Cherenkov effect. 
The latter gives an unusual insight into the instability in modified gravity from the perspective of common physics.

\section{Cherenkov radiation in 2D and higher dimensions}

For simplicity we will first focus on the case of 1+1 dimensions and then discuss cases of higher dimension.
In this section we consider basic properties of wave propagation that are important to describe the Cherenkov radiation in a convenient  way in the following discussion.
Throughout the paper we assume flat spacetime with non-dynamical metric.

A massless scalar field in 1+1 dimensions with the speed of propagation $c_s$ obeys the equation
\begin{equation}
\label{eq_phi}
\ddot\phi-c_s^2\phi'' =0.
\end{equation}
The above equation is written in coordinates where the cone of propagation is symmetric. 
By doing any Lorentz boost with the speed different from $c_s$ (in particular, with the speed of light $c=1$), Eq.~(\ref{eq_phi}) takes a more complicated form, see e.g.~\cite{Babichev:2007dw}. 
The solution of~(\ref{eq_phi}) can be presented in terms of Fourier modes, 
\begin{equation}
\label{sol_phi}
\phi = A_k e^{-i\omega t+ikx},
\end{equation}
such that
\begin{equation}
\label{omegak}
\omega = \pm c_s k.
\end{equation}
The above equation is the simplest dispersion relation, it relates the frequency of the wave $\omega$ to the momentum $k$.
Usually $\omega$ is associated to the energy, so that in Eq.~(\ref{omegak}) one assumes positive $\omega$, while $k$ runs from $-\infty$ to $+\infty$.
The wave~(\ref{sol_phi}) propagates along the constant phase $\Phi \equiv -\omega t+kx=\text{const}$, which also corresponds to characteristics of Eq.~(\ref{eq_phi}). 

It is worthwhile to make a distinction between the propagation vector $N^\mu$ along which the wave propagates, and the momentum $k^\mu$ corresponding to the massless particle in this model. 
Indeed, the momentum is found as the gradient of the phase,
\begin{equation}
\label{k}
k_\mu = \nabla_\mu\Phi = \{-\omega, k\} \quad \rightarrow \quad k^\mu = \{\omega,k\}
\end{equation}
where in the last relation one can easily recognize the standard expression for the momentum of a particle. 
The energy of the particle is given by $\omega$, while $k$ gives the spatial momentum.
If the particle is subluminal, then we have $\omega<k$. In particular for the case of dust $\omega=0$.  
For the particles propagating with the speed of light $\omega=k$. In the case of superluminal particles one can see that $\omega>k$.

On the other hand, the propagation vector $N^\mu$ is found from the expression $N^\mu k_\mu =0$, since it is tangential to the surface $\Phi=\text{const}$.
In the case under consideration, one easily finds that 
\begin{equation}
\label{N}
	N^\mu = \{k,\omega\}.
\end{equation}
Comparing~(\ref{k}) and~(\ref{N}) it is clear that $N^\mu$ and $k^\mu$ are in general two different vectors. 
An important object for us in what follows is the wave vector $k^\mu$ for different species of particles, rather than the propagation vector, 
since it is $k^\mu$ which is engaged in conservation of energy-momentum, the key notion for both Cherenkov radiation and ghost instability. 
However, when necessary, we will also make a reference to the propagation cones, in order to keep the discussion clear. 

Let us take a look at the Cherenkov radiation in the simplest case of 1+1 dimensions. 
We consider medium at rest with a subluminal speed of perturbations $c_s<1$ and we will refer to particles corresponding to perturbations of the medium as to phonons. 
We also take one massive particle with the standard dispersion relation 
\begin{equation*}
\omega^2 = m^2 + k^2,
\end{equation*}
where $m$ is the mass of the particle, that moves through the medium. 
Provided that there is an interaction between the particle and the medium (or, rather, with phonons), the moving particle may transfer energy to phonons, if the energy-momentum conservation is satisfied.
For simplicity we assume that the energy of the Cherenkov particle is much larger than the emitted phonon energy, so that we can neglect the change in the energy of the moving particle when a photon is emitted.

\begin{figure}[t]
\includegraphics[width=0.9\textwidth]{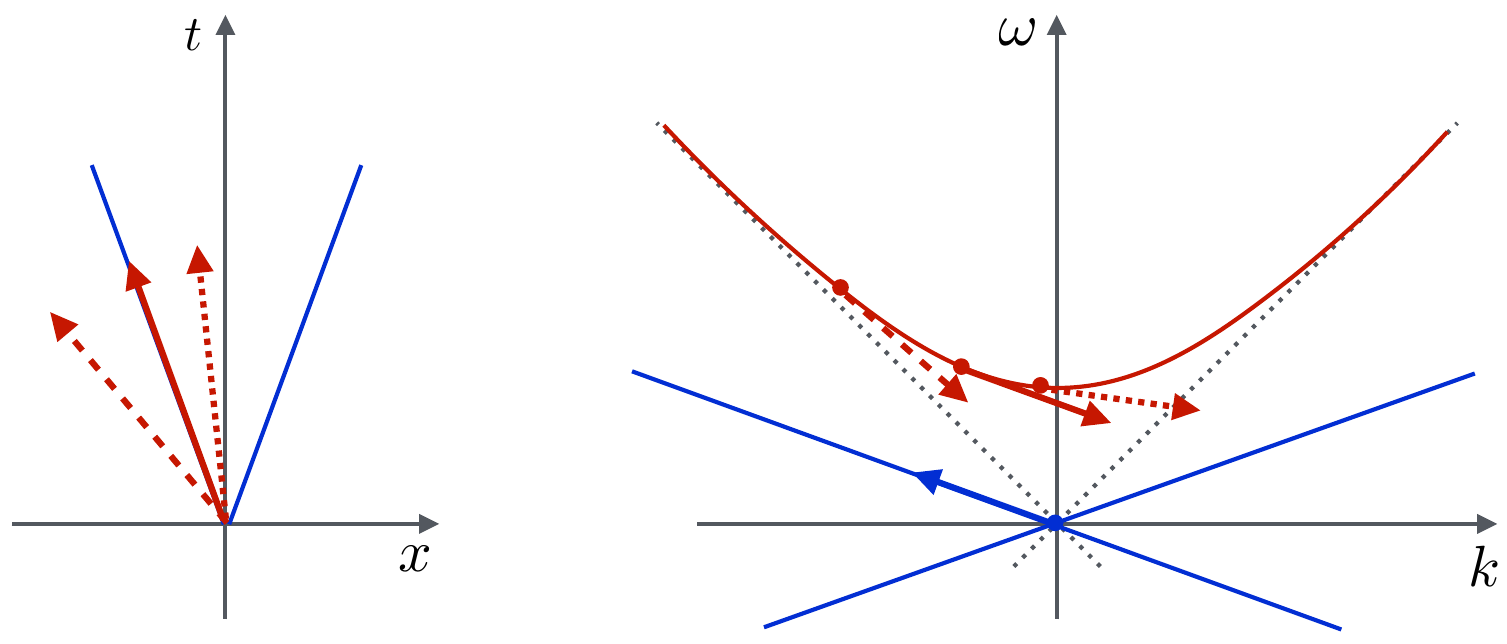}}{\caption{On the left plot the cone of propagation (blue) of the perturbations of the medium, 
and the momentum vector of propagation of a particle propagating at a subsonic (dotted red), sonic (solid red) and supersonic (dashed red) speed, are shown. 
The corresponding dispersion relations are depicted on the right plot. 
When the velocity of the particle becomes equal to the speed of phonons, Cherenkov radiation occurs. The particle looses its energy and momentum (red solid arrow) by creating a phonon (blue arrow).}
\label{fig_cherenkov1}
\end{figure}

The Cherenkov radiation occurs when the speed of the particle becomes equal to the speed of phonons, see Fig.~\ref{fig_cherenkov1}. 
Indeed, it is not difficult to see that the particle can lose energy to a phonon with some probability; 
and the energy-momentum conservation is satisfied as long as vector tangential to the curve $\omega=\omega(k) = \pm\sqrt{m^2+k^2}$ of the particle is parallel to the straight  line $\omega = \pm c_s k$ corresponding to the dispersion relation of phonons. 
This happens when the velocity of the particle is equal to the speed of perturbations.
Note that here we neglected the backreaction of the emitted radiation on the moving particle, as discussed above.
If the particle has the momentum states $p_1^\mu$ and $p_2^\mu$ before and after it loses energy to the phonon, then the following relation holds,
\begin{equation*}
p_1^\mu =  p_2^\mu + k^\mu,
\end{equation*}
which we rewrite more conveniently as
\begin{equation}
\label{cherenkov}
0=\Delta p^\mu + k^\mu,
\end{equation}
where $k^\mu$ is the momentum of the phonon and $\Delta p^\mu= p_2^\mu - p_1^\mu$.

If the particle has even higher velocity than the phonon speed, the standard Cherenkov radiation does not take place in 2 dimensions. 
This happens due to the kinematic reason: the momentum cannot be conserved in this case, as it can be seen from Fig.~\ref{fig_cherenkov1}: 
$-\Delta p^\mu$ cannot be equal to the phonon momentum $k^\mu$, meaning that the process with emitting one phonon by the Cherenkov particle is impossible, i.e. Eq.~(\ref{cherenkov}) cannot be satisfied.
Instead, as we discuss later, emission of two phonons is possible, although this process is suppressed as compared to the standard Cherenkov radiation. 

\begin{figure}[t]
\includegraphics[width=0.5\textwidth]{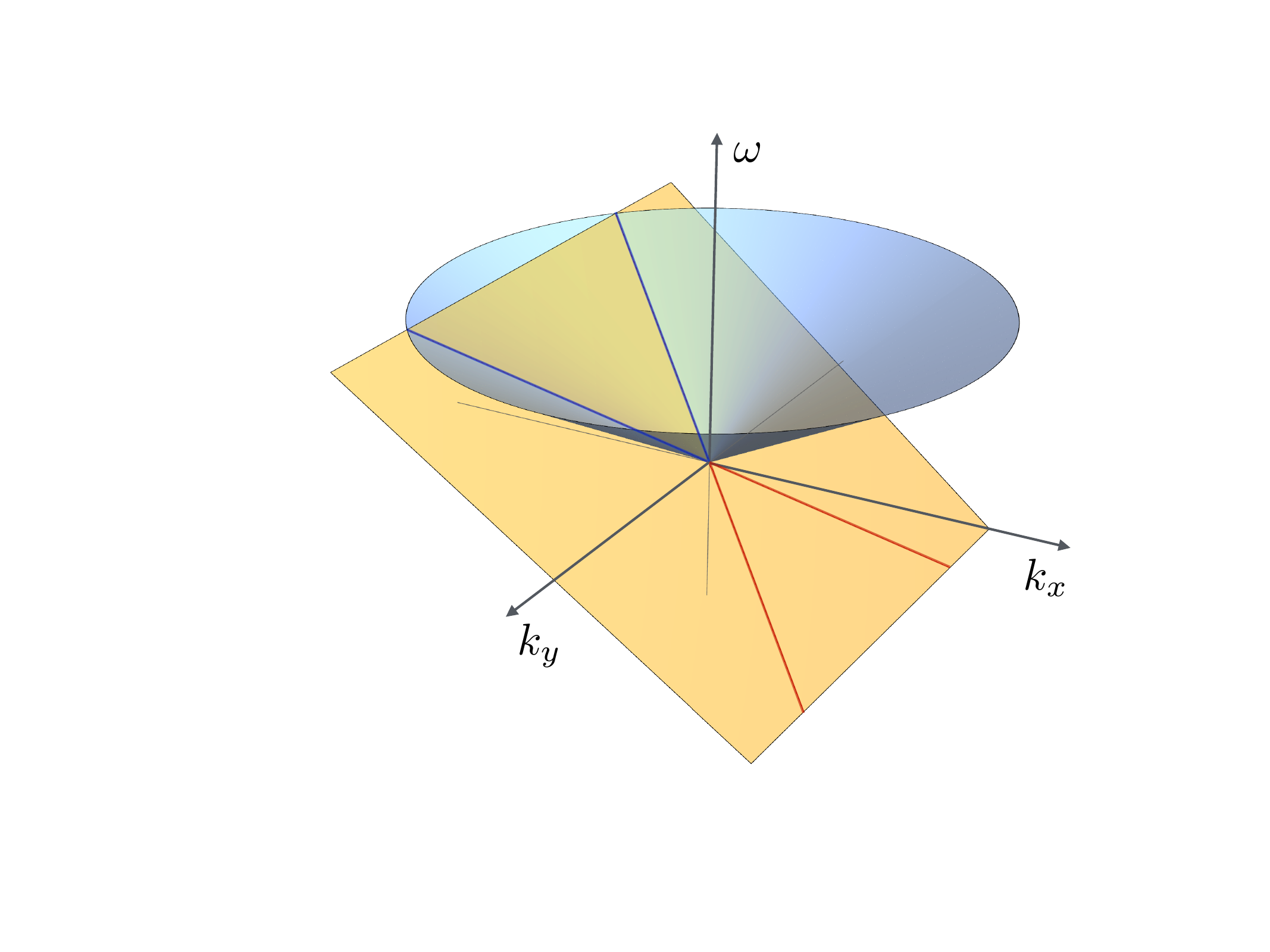}}{\caption{The dispersion relation for the phonons of the medium, shown by the cone, and the tangential plane to the mass surface of the massive particle, moving through the medium at the speed higher than the speed of perturbations, in 2+1 dimensions. 
The intersection of the two surfaces, shown by blue lines, corresponds to the kinematically allowed values of the momenta and energy of emitted phonons. 
The red lines indicate the corresponding change in momentum of the moving particle.
The projection of blue lines to the $\{k_x,k_y\}$ plane gives the direction of propagation of Cherenkov radiation, which corresponds to the Cherenkov angle.}
\label{fig_cherenkov2}
\end{figure}

The kinematic constraint for the Cherenkov particle moving at the speed higher than the speed of perturbations in 1+1 is evaded if another spatial dimension (or more) is added. 
The Cherenkov process then takes place in a standard way with the presence of the Cherenkov cone of radiation~\cite{ginzburg,jackson}.
To find the Cherenkov cone, we need to replace the right plot of Fig.~\ref{fig_cherenkov1} by a 3D picture where two components of the spatial part of momentum are present, $k_x$ and $k_y$, see Fig.~\ref{fig_cherenkov2}.
The dispersion relation for phonons is represented by the cone in Fig.~\ref{fig_cherenkov2},
\begin{equation*}
\omega^2 = c_s^2\left(k_x^2+ k_y^2\right),
\end{equation*}
while the dispersion relation for the Cherenkov particle becomes the mass surface (not shown in Fig.~\ref{fig_cherenkov2}), 
\begin{equation*}
\omega^2 = m^2 + k_x^2+ k_y^2.
\end{equation*}
We assume that a particle propagates along $x$ direction, so the change in its momentum is given by
\begin{equation}
\label{Delta}
\Delta p^\mu = \{v\Delta p_x, \Delta p_x, \Delta p_y \},
\end{equation}
where $v$ is the velocity of the particle and we again assumed that the particle is heavy enough to neglect the backreaction of the emitted radiation. 
The vector $\Delta p^\mu$ of the change of momentum of a moving particle is located on the surface that is tangential to the mass surface, it corresponds to the plane in Fig.~\ref{fig_cherenkov2}.
The emitted phonon has the momentum 
\begin{equation}
\label{phonons2}
k^\mu = \{\pm c_s\sqrt{k_x^2+k_y^2}, k_x, k_y \}
\end{equation}
and are shown by the cone in Fig.~\ref{fig_cherenkov2}. 
By conservation of energy-momentum, one arrives at the well-known expression for the Cherenkov cone:
\begin{equation}
\label{angle}
\cos\theta = \frac{c_s}{v}.
\end{equation}
This is a standard result for Cherenkov radiation~\cite{ginzburg,jackson}. 
Graphically in Fig.~\ref{fig_cherenkov2}, the conservation of energy-momentum means that there is an intersection (shown by red and blue lines) of the two surfaces --- the plane, corresponding to the change in the momentum of the moving particle $\Delta p^\mu$, and the cone of dispersion relation for propagating phonons~(\ref{phonons2}).

It is also clear from~(\ref{angle}) that in case of 1+1 dimension, the velocity of the particle must be exactly equal to the speed of the perturbations of the medium  in order for the Cherenkov effect to happen.
As we discussed in the above, due to kinematic reasons, the Cherenkov radiation is forbidden for particles traveling with higher speeds in this case, and one needs to add another dimension, so that the Cherenkov radiation would be kinematically allowed. 
However, if we remain in 1+1 dimensions, is it still possible to have an analogue of the Cherenkov radiation in a multiple-particle process. 
For example, a process with two created phonons is possible:
\begin{equation}
\label{cherenkov2}
	0 = \Delta p^\mu + k^\mu + \tilde{k}^\mu.
\end{equation}
Indeed, using~(\ref{Delta}) and (\ref{phonons2}), one can find the the energy-momentum conservation (\ref{cherenkov2}) is satisfied provided that
\begin{equation}
\label{sol2}
\Delta p^\mu = \{-v, -1 \}\Delta p, \;\; k^\mu = \frac{v-c_s}{2}\{ c_s, -1 \}\Delta p , \;\; \tilde{k}^\mu =\frac{v+c_s}{2}\{ c_s, 1 \}\Delta p.
\end{equation}
In case of higher dimensions this process is subdominant compared to the standard Cherenkov radiation, however, in 1+1 dimension, when the standard Cherenkov radiation is absent for $v>c_s$,
the emission of two phonons~(\ref{cherenkov2}) is the dominant contribution.

\section{Ghosts and vacuum instability}
\label{sec_ghost}
\begin{figure}[t]
\includegraphics[width=0.95\textwidth]{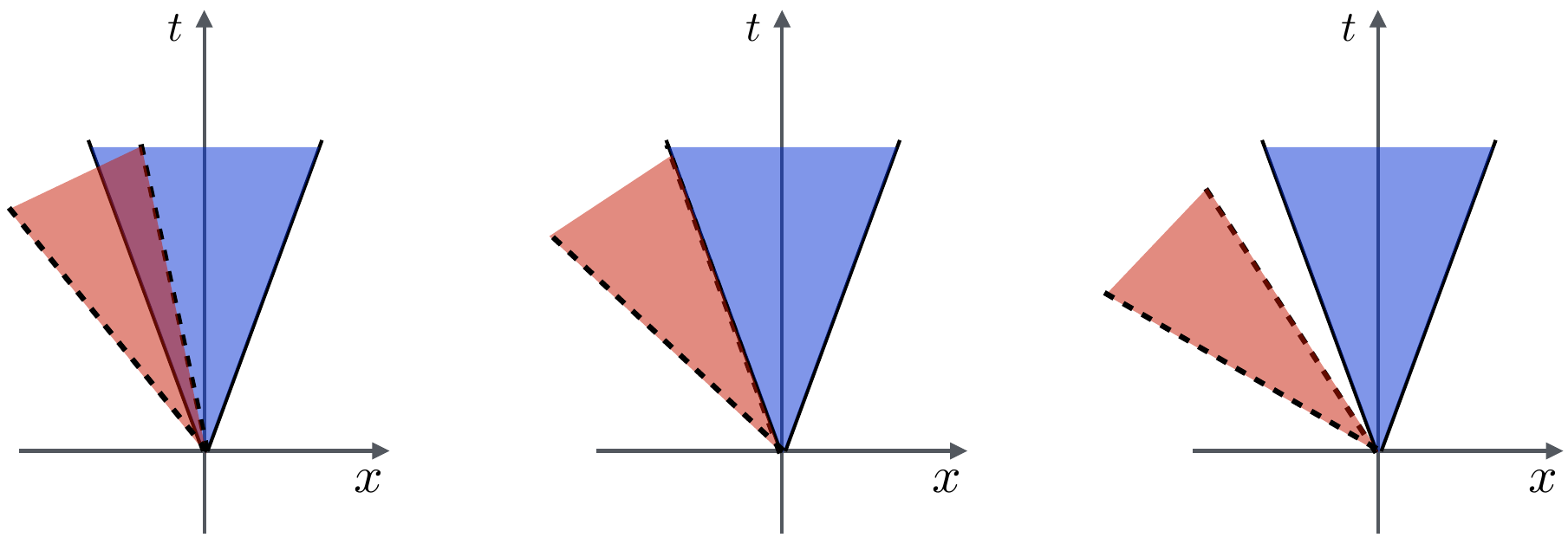}}{\caption{Three different mutual configurations of cones are shown. On the left plot the cones have a common Cauchy surface as well as common time, therefore there is no vacuum decay. On the right plot the cones do not have a common time, therefore the energy of the system of two fields is not positive definite, so that vacuum decay is possible.
The middle plot is the marginal case, when the cones just touch each others from outside. In this case the vacuum decay just becomes possible.
}\label{fig_cones}
\end{figure}

Now we turn to the question of ghosts and associated vacuum instability. 
We have in mind a species which energy is negative is some coordinates, while some other species has a positive energy in the same coordinates.
The full analysis of all possible scenarios have been treated in~\cite{Babichev:2018uiw}, where the criteria for the system of two (or more) species in 1+1 dimensions to be stable have been presented.
Here we will be interested in particular case of~\cite{Babichev:2018uiw}, when two cones of different species have a common Cauchy surface, but they do not have a common time, as it is depicted in the right plot of  Fig.~\ref{fig_cones}. 
As it was shown in~\cite{Babichev:2018uiw}, the species corresponding to the red cone of propagation have negative energy, while the blue particles carry positive energy. 
Therefore the vacuum decay may become possible. 
Indeed, one may expect creation of particle with positive energy and a particle of negative energy from nothing, thus the vacuum destabilizes, provided that the process is kinematically allowed. 
Therefore a solution, corresponding to the middle or right panel of Fig.~\ref{fig_cones} should suffer from instability.

\begin{figure}[t]
\includegraphics[width=0.95\textwidth]{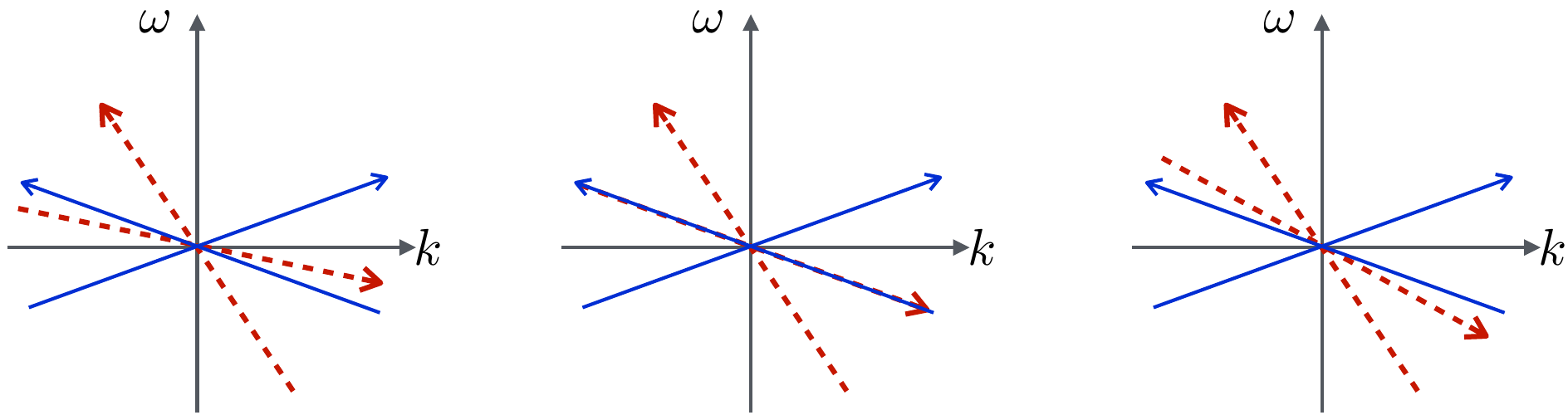}}{\caption{The dispersion relations corresponding to the situations depicted in Fig.~\ref{fig_cones}.
On the left figure, the energy of the red particle can be negative, however, the vacuum decay is kinematically forbidden. On the right plot the vacuum decay is possible.
The middle plot is the marginal case, when the vacuum decay starts.
}\label{fig_dispersion}
\end{figure}

Here we discuss the instability of such a system suggested in~\cite{Babichev:2018uiw} and consider in detail creation of a pair of particles with kinematics of the process taken into account. 
We also relate this process of vacuum decay the Cherenkov radiation described in the previous section.

Let us first consider the marginal case when the two cones touch, in 1+1 dimensions, the situation depicted in the middle plot of Fig.~\ref{fig_cones}. 
In order to check whether the vacuum decay is possible, we draw the dispersion relation corresponding to propagation of the two cones, see Fig.~\ref{fig_dispersion}.
It is clear from the middle plot of Fig.~\ref{fig_dispersion}, that the creation of a couple of two particles, corresponding to two different species, is possible,
\begin{equation}
\label{decay}
0= p^\mu +k^\mu ,
\end{equation}
where $k^\mu$ and  $p^\mu$  are the particles corresponding to the blue and red cones respectively. 

Now, comparing the Cherenkov radiation, shown on the right plot of Fig.~\ref{fig_cherenkov1}, and the vacuum decay demonstrated in the middle plot of Fig.~\ref{fig_dispersion}, we can see that kinematic pictures of these two processes are fully equivalent.
In both cases phonons with positive energy (blue particles) are created when it becomes kinematically allowed for the negative energy to be compensated with: 
in the case of Cherenkov radiation it is a difference between the energies of the two states of a moving particle $\Delta p^\mu$, while in the case of vacuum decay it is emergence of a ghost with wave vector $p^\mu$. 

\begin{figure}[t]
\includegraphics[width=0.5\textwidth]{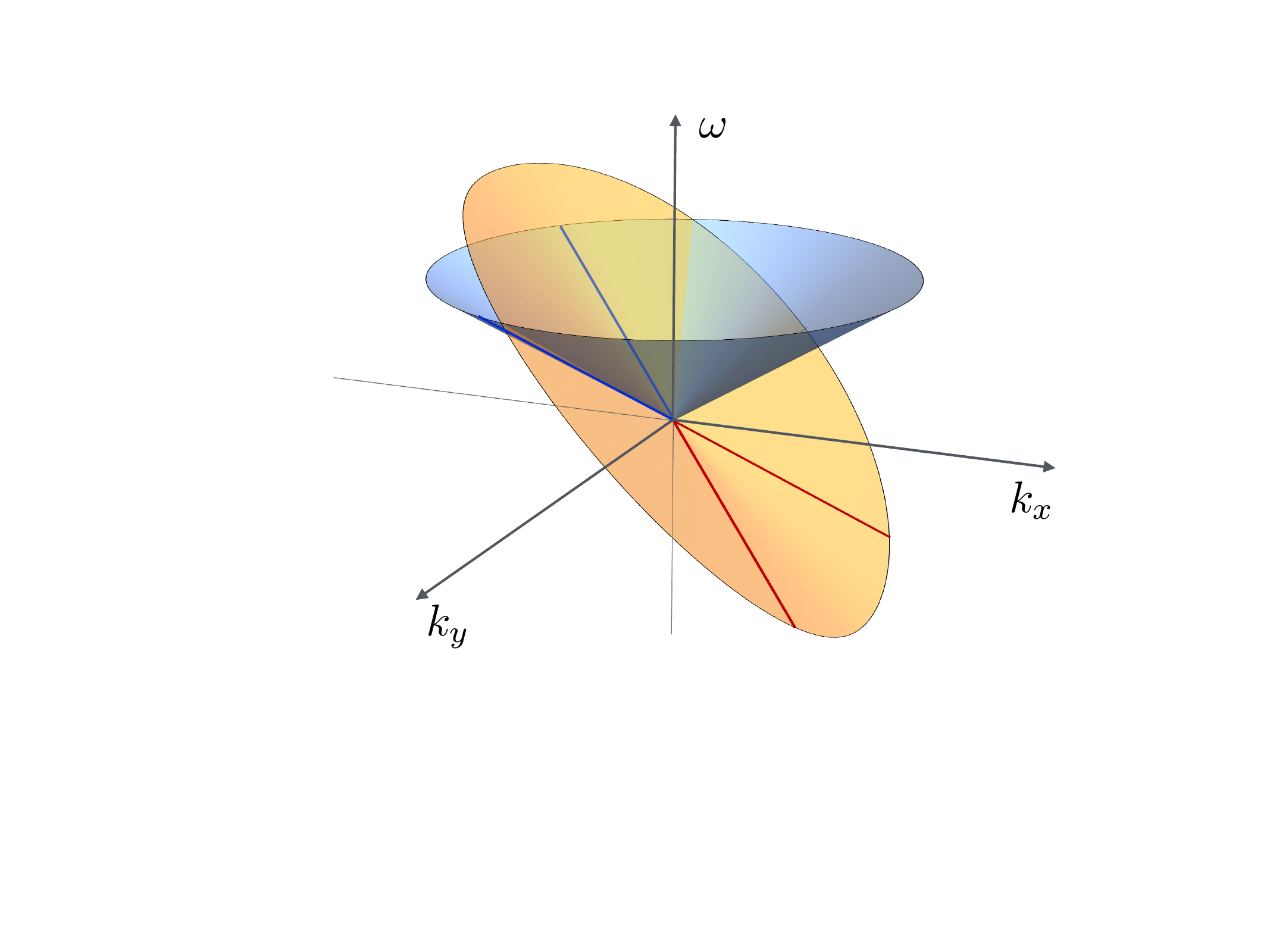}}{\caption{The cones corresponding to the dispersion relations of the two species, a 2+1 analogue of the right plot of the Fig.~\ref{fig_dispersion}.
The particle with positive energy is represented by the blue cone, while the ghost particle is shown by orange.
The intersection of the ghost cone with the \emph{past} photon cone, shown by red lines, corresponds to the kinematically allowed values of the momenta of created ghost particles. 
The production of the ghost particles is accompanied by creation of phonons, shown by blue lines on the blue \emph{future} cone.
The projection of these lines to the $\{k_x,k_y\}$ plane gives the direction of propagation of created phonons and ghost particles, which is similar to the Cherenkov cones of radiation.
}
\label{fig_ghost2}
\end{figure}

In fact this analogue goes for all the aspects of the Cherenkov radiation that we considered in the previous section. 
Indeed, similar to the case of the Cherenkov radiation in 2D, the vacuum decay involving only 2 particles---one normal particle and one ghost---is only possible in the marginal case, when the cones touch each other from outside. 
This is equivalent to the Cherenkov radiation of a particle moving with the velocity equal to the speed of propagation of phonons, as we discussed above.
When the cones are separated, as it is shown on the right plot of Fig.~\ref{fig_cones}, the vacuum decay as described above is  kinematically forbidden, similar to the case of Cherenkov radiation, when the particle moves at velocity higher than the speed of phonons.
However, when an extra dimension is added, exactly like in the case of Cherenkov radiation, the vacuum decay into two particles becomes again possible.
Indeed, in 2+1 dimensions, one needs to extend the right plot of Fig.~\ref{fig_dispersion} to include an extra dimension, and to consider a 3D picture in coordinates $(k_x, k_y, \omega)$.
The dispersion relations are then graphically represented by cones, see Fig.~\ref{fig_ghost2}. 
Then the kinematically allowed values of the momenta and frequency will correspond to intersection of these two cones in the momenta space. 
Such an intersection always takes place when the cones of propagations are separated, i.e. the situation shown on the right plot of Fig.~\ref{fig_cones}, 
which automatically implies the existence of lines of intersection as it is depicted in the Fig.~\ref{fig_ghost2} in the momenta space.
In this case the products of vacuum decay form two cones with opposite momenta: one being composed of normal particles and the other made of ghost particles. 
This is similar to Cherenkov radiation, with the only exception that there are now two cones instead of one Cherenkov cone.

If we however stick to 1+1 dimension, the vacuum decay into two particles is forbidden in case of separated cones, see the right plot of Fig.~\ref{fig_cones}. The `decay cones' do not exist in this case, similar to the non-existence of the Cherenkov cone for very high velocity of the particle. 
This can be easily understood from the right plot of Fig.~\ref{fig_dispersion}: there is no intersection of dispersion relation cones in this case.
Instead, one can consider a multi-particle decay.
To keep a close analogue to the Cherenkov case, we consider a creation of one ghost particle (corresponding to the red cones in Figs.~\ref{fig_cones} and \ref{fig_dispersion}) and two normal (blue) particles:
\begin{equation}
\label{ghost2}
	0 = p^\mu + k^\mu + \tilde{k}^\mu.
\end{equation}
For this process, the technical derivation is fully equivalent to the process of the multiple particle ``Cherenkov'' radiation, considered in the previous section, see Eq.~(\ref{cherenkov2}). 
One simply needs to replace $\Delta p^\mu$ of the Cherenkov case to $p^\mu$ in the case of vacuum decay. 
Two normal particles created in the process are equivalent to two Cherenkov phonons, while the ghost particle that emerge from the vacuum decay, Eq.~(\ref{ghost2}), is equivalent to the difference of the momenta $\Delta p^\mu$ of the Cherenkov particle.

A natural question then can be posed about the dynamics of the process. 
The Cherenkov radiation is a real process, which has been observed in many physical situations. 
On the other hand, vacuum decay due to the presence of a ghost is considered to be a dangerous instability that sometimes thought to be instantaneous. 
There is no contradiction, however,  if one takes into account details of the Cherenkov radiation. 
In physical situations Cherenkov process is not instantaneous, i.e. a Cherenkov particle does not stop momentarily, due to the physical cutoff for the process.
Indeed, the phonons cease to exist for very high energies, when the cutoff of the momentum is at least of order of the inverse distance between atoms. 
In case of photons, the dispersion relation in medium is momentum-dependent so that the phase velocity of photons grows as $k$ increases. 
This ensures that the energy loss of the Cherenkov particle remains finite. 
Indeed, according to Frank-Tamm formula~\cite{Frank:1937fk}, the energy loss per unit time reads~\cite{ginzburg,jackson,nair},
\begin{equation}
\label{FT}
\frac{d E}{d t}= e^2 v \int\left(1-\frac{c^2_p(\omega)}{v^2 }\right) \omega d \omega,
\end{equation}
where $e$ is the charge of the electron, $v$ is the speed of the moving particle, and the phase velocity of the photons in the media $c_p$  plays the same role as $c_s$ in our discussion.
In the high-frequency limit, the phase velocity tends toward the speed of light in vacuum, which guarantees the finiteness of the integral in~(\ref{FT}).
(Note that the domain of integration is restricted to those frequencies $\omega$ for which $c_p(\omega)/v \leq 1$).

A similar consideration concerning a cutoff applies for the process of  vacuum decay as described in Sec.~\ref{sec_ghost}. 
Indeed, if a modified gravity theory is considered to be an effective field theory, then there is a cutoff of the theory, beyond which the description in terms of a specific modified gravity model becomes invalid. 
The cutoff and the background solution determine the largest momenta of created pairs of normal particles and ghosts. This in turn yields a finite rate of a ghost instability.

Let us now examine in more detail the dynamics of vacuum decay\footnote{I thank Sabir Ramazanov for helpful discussions on this matter.}. 
To make the comparison with the standard Cherenkov radiation more transparent, we will switch to four dimensions for the remainder of this section.
As discussed above, the dominant vacuum decay process is into two particles---one normal and one ghost, see Eq.~(\ref{decay}).
The decay rate can be obtained by a straightforward modification of the standard particle decay expression, given in e.g.~\cite{Peskin:1995ev},
\begin{equation}
\label{rate}
\Gamma_{\mathrm{vac} \rightarrow p+k}=\int \frac{d^3 p}{(2 \pi)^3} \frac{1}{2 \omega_p}\; \frac{d^3 k}{(2 \pi)^3} \frac{1}{2 \omega_k}\left|\mathcal{M}\left(\mathrm{vac} \rightarrow p+k\right)\right|^2(2 \pi)^4 \delta^{(3)}\left(\vec{p}+\vec{k}\right)\delta\left(\omega_p+\omega_k\right),
\end{equation}
where $\mathcal{M}\left(\mathrm{vac} \rightarrow p+k\right)$ is the matrix element corresponding to the process~(\ref{decay}) and it depends on the details of interaction between a normal particle $k$ and a ghost $p$ (See also similar discussions in a different context  in~\cite{Sbisa:2014pzo,Konnig:2016idp,Ramazanov:2016xhp}).
Integrating in~(\ref{rate}) over $\vec{k}$ and then going to the polar coordinates $d^3 p = (2\pi)p^2\; dp\; \sin\theta d\theta$, where $p=|\vec{p}|$ and the axis of symmetry is aligned with the tilt of one of the cones (e.g. $k_x$ in Fig.~\ref{fig_ghost2}), one finds,
\begin{equation}
\label{rate1}
\Gamma_{\mathrm{vac} \rightarrow p+k}=\frac{1}{8\pi} \int \frac{p^2 d p \sin\theta d\theta}{\omega_k \omega_p} \left|\mathcal{M}\left(\mathrm{vac} \rightarrow p+k\right)\right|^2\delta\left(\omega_p+\omega_k\right).
\end{equation}
To take into account the delta-function in the latter expression, one needs the expressions for the dispersion relation of the normal and ghost particles, $\omega_k(k)$ and $\omega_p(p)$ correspondingly. 
For non-ghost particle (blue cone in Fig.~\ref{fig_ghost2}) one can write
\begin{equation}
\label{dispersion1}
\omega_k = c_s p,
\end{equation}
where we took into account that $\vec{k}=-\vec{p}$.
For the ghost particle the dispersion relation is more complicated. 
In the frame where the cone is symmetric, the dispersion relation is analogous to that of the normal particle, $\omega_p^{\prime 2} = \tilde{c}_s^2\left(p_x^{\prime 2}+ p_y^{\prime 2}+ p_z^{\prime 2}\right)$, where $\tilde{c}_s^2$ is the speed of propagation in this frame.
In a frame boosted with velocity $v$ along the $x$-direction, the dispersion relation takes the form
\begin{equation}
\label{dispersion2}
\frac{\left(\omega_p+v p \cos\theta\right)^2}{1-v^2}=\tilde{c}_s^2\left[\frac{\left(p\cos\theta+v \omega_p\right)^2}{1-v^2}+p^2\sin^2\theta\right].
\end{equation}
Then solving~(\ref{dispersion2}) for $\omega_p$ and substituting the obtained expression along with~(\ref{dispersion1}) in the equation $\omega_k+\omega_p=0$, 
one can find the `decay angle' $\theta_0$ in terms of $c_s$, $\tilde{c}_s$ and  $v$. 
Note that $p$ drops out from the final expression. This allows to integrate over $\theta$ in~(\ref{rate1}), however, the expression for $\theta_0$ is rather complicated.

For the sake of simplicity, let us take the limit $\tilde{c}_s\to 0$, corresponding to the dust-like behavior of the second (ghost) particle. 
In this case the dispersion relation cone becomes effectively a plane and  one  arrive at the situation, best described by Fig.~\ref{fig_cherenkov2}. 
Assuming $\tilde{c}_s\to 0$ in~(\ref{dispersion2}) gives a simple expression $\omega_p = - v p \cos\theta$.
The delta-function in~(\ref{rate1}) reads 
$$\delta(c_s p - vp\cos\theta )$$ 
which ensures that the vacuum decay is only possible in the angle $\cos\theta=\cos\theta_0$, where $\cos\theta_0 = c_s/v$. 
In the latter expression for $\theta_0$ one recognizes the Cherenkov angle. 
Ghost particles $p$ emerge at an angle $\theta_0$ relative to the $x$-axes, whereas normal particles are emitted oppositely, at angle $\pi -\theta_0$ (see also Fig.~\ref{fig_ghost2}). 
Finally, making an integration over $\theta$ in Eq.~(\ref{rate1}) we arrive at the vacuum decay rate,
\begin{equation}
\label{rate2}
\Gamma_{\mathrm{vac} \rightarrow p+k}=\frac{1}{8\pi v c_s^2} \int  \left|\mathcal{M}\left(\mathrm{vac} \rightarrow p+k\right)\right|^2 \frac{d\omega}\omega.
\end{equation}
The matrix element $\mathcal{M}$ in Eq.~(\ref{rate2}) depends on the details of the interaction between the two species.
In general, it may depend on $\theta$, in this case in the expression~(\ref{rate2}) it should be taken at $\theta=\theta_0$. 
$\mathcal{M}$ may as well depend on the energy $\omega$, similar to what happens in the case of Cherenkov radiation. 
In case when it is independent on $\omega$, the integral in~(\ref{rate}) diverges logarithmically. A cutoff, which we discussed above, renders the integral finite in the UV, independently on the behavior of $\mathcal{M}$.

Note that apart from the cutoff of the effective field theory, there is yet another limitation for the rate of a ghost instability, which applies to majority of modified gravity theories. 
Similar to the Cherenkov radiation, a situation of two different species having different cone orientation, as shown in Fig.~\ref{fig_cones}, 
can only occur if at least one of the species corresponds to perturbations on a non-trivial background. 
This is because the theories we have in mind are in fact Lorentz invariant, while spontaneous breaking of Lorentz invariance happens only on non-trivial background.
The background is  eventually destroyed by the creation of more and more perturbations on top of it, therefore the process of decay may stop at some moment.  
This is again very similar to Cherenkov radiation: eventually all Cherenkov particles slow down and stop emitting radiation.


\section{Discussion}

In this paper we discussed the relation between standard Cherenkov radiation arising when a particle moves in a medium at the velocity higher than the speed of propagation in this medium, 
and ghost instability that is common  in modified gravity theories.

In the simplest case of 1+1 dimensions, the conventional Cherenkov radiation occurs when the velocity of a particle is equal to the speed of phonons of the medium, see Fig.~\ref{fig_cherenkov1}. 
In this case it is kinematically allowed that the Cherenkov particle loses its energy and momentum by emitting one phonon. 
This process is fully equivalent to the vacuum decay in 2D, when the cones of two species touch each others from outside, see the middle plots of Figs.~\ref{fig_cones} and \ref{fig_dispersion}.
On the level of equations the two processes are the same. 
The physical interpretation, however, may naively seem to be different. 
Indeed, in case of Cherenkov radiation, a moving particle loses its energy by transferring it to perturbations of the medium. 
For the vacuum decay, as it rather often occurs in gravity theories, the two particles are born from nothing, one of those is a normal particle with positive energy, while the other is a ghost particle carrying negative energy.
In order to see why for these physical situations the underlying mechanism is the same, one may think of a moving Cherenkov particle as a part of a collection of many particles that form dust, i.e. medium with a zero speed of propagation of perturbations. 
Then, if one assumes the initial energy of the dust configuration to be zero, the change in energy of each particle due to the Cherenkov emission can be viewed as appearance of a perturbation with negative energy on top of the dust background. 
The cone of propagation of these ghost-like Cherenkov perturbations shrinks to a line, and Cherenkov radiation can be interpreted in terms of ghost instability by identifying the red cone in Fig.~\ref{fig_cones} (which is infinitely thin in this case) with the change of energy-momentum of Cherenkov particles.
Thus the Cherenkov process is fully analogous to the vacuum decays as it has been described in Sec.~\ref{sec_ghost}.

This correspondence between Cherenkov and ghost instability goes further.
In 2D case, strictly speaking, the conventional Cherenkov radiation is absent for a particle moving faster than the speed of phonons. 
The reason is purely kinematic: the energy-momentum conservation does not allow for such a process, Fig.~\ref{fig_cherenkov1}. 
The same is true for the vacuum decay, when the cones of propagation are separated, the right plot of Fig.~\ref{fig_cones}, the vacuum decay into two particles is forbidden for the same kinematic reason.
However, a multiple-particle vacuum decay is possible instead, in particular, creation of one ghost particle and two normal particles is allowed, Eq.~(\ref{ghost2}). 
Therefore vacuum decays by multiple particle creation in this case. 
Analogously, a multiple-particle version of Cherenkov radiation is possible, see Eqs.~(\ref{cherenkov2}) and (\ref{sol2}).
These two effects---multiple-particle Cherenkov radiation and the multiple-particle vacuum decay in the presence of a ghost---are fully equivalent. 
To see this more clearly, one should again take a point of view that the Cherenkov particles form a cloud of dust with zero energy of perturbations, and when the Cherenkov emission occurs, negative perturbations on top of background dust solution appear. This process is are fully equivalent to a ghost particle creation, considered in Sec.~\ref{sec_ghost}.

To make the analogy between two physical processes even more compelling, we considered a slightly more complicated case of 2+1 dimensions.
In 2+1 dimensions, when a particle moves in the medium faster than the speed of propagation in the medium, the Cherenkov cone appears. 
I.e. the moving particle transfer its energy to phonons have a particular emission angle with respect to the moving particle, Eq.~(\ref{angle}).
It is the presence of an extra spatial dimension this allows the phonons to be emitted, while   the kinematics defines the angle of emission, the Cherenkov cone.
By examining the vacuum decay in 2+1 dimensions, one arrives to a similar result: normal and ghost particles are created by vacuum decay with a certain `decay' angle.
In this regards, again, the Cherenkov radiation and the vacuum decay are two analogous processes.

As we have shown, the kinematics of the two processes---Cherenkov radiation and the ghost instability of a certain type---are indeed fully equivalent. 
One the other hand, their dynamics also reveals similar features, such as the existence of a physical cutoff in both theories, as discussed in Sec.~\ref{sec_ghost}. 
We calculated vacuum decay rate, Eq.~(\ref{rate2}), which depends on details of interaction between the normal and ghost particle, via the matrix element. 

Besides the UV cutoff, ghost instabilities are further constrained because distinct cone orientations, Fig.~\ref{fig_cones}, can only arise on non-trivial backgrounds (for Lorentz-invariant theories) where Lorentz invariance is spontaneously broken. 
As particle production depletes the background, the decay process may naturally cease, much like Cherenkov radiation ends once particles slow down below the emission threshold.

To conclude on vacuum decay considered in this paper, the presence of this type of ghost renders solution in a theory of gravity to be unstable, in accordance with Ref.~\cite{Babichev:2018uiw}. 
This decay is analogous to the instability of Cherenkov particles that loose energy due to emission of radiation. 
This instability is not instantaneous: the time of instability is determined by the cutoff of the theory under consideration.

It would be interesting to study a scenario, when such an unstable configuration in modified gravity is quasi-stable, that is the time of instability being much longer than relevant physical processes. 
For example, one can think of a black hole with the presence of a ghost, but with instability rate smaller than the frequency of quasinormal modes.
In addition, from observational perspective, the emission of normal particles and ghosts with particular `ghost cone' direction, discussed  in Sec.~\ref{sec_ghost}, might be worth investigating further.
Another direction for future study would be to examine analytically and numerically how such type of ghost instability is developed for particular solutions in various gravity theories.
\\

\emph{Note added.} The paper~\cite{Sawicki:2024ryt}, which explores a topic partially overlapping with this study, appeared simultaneously with our work.

\section*{Acknowledgements}
I would like to thank Alessandro Fabbri, Gilles Esposito-Far\`ese, Sabir Ramazanov, Ignacy Sawicki and Alexander Vikman for interesting and stimulating discussions.
The work was supported by ANR grant StronG (ANR-22-CE31-0015-01).


\end{document}